\documentclass[
aps,
prl,
preprint,
superscriptaddress,
amsmath,
amssymb,
showpacs
]{revtex4-1}

\usepackage[utf8x]{inputenc}

\usepackage{color}
\usepackage{amsmath}
\usepackage{amssymb}
\usepackage{amsfonts}
\usepackage{amsthm}
\usepackage{float}
\usepackage{bm}
\usepackage{latexsym}
\usepackage{hyperref}
\usepackage{multirow}
\usepackage[capitalize]{cleveref}
\usepackage{graphicx}
\hypersetup{
    colorlinks,
    citecolor=blue,    filecolor=blue,
    linkcolor=blue,     urlcolor=blue
}

\begin{document}


\title{Hard X-rays from laser-wakefield accelerators in density tailored plasmas}

\author{Michaela Kozlova}
\affiliation{ELI Beamlines, Institute of Physics CAS, Prague, 182 21, Czech Republic}
\affiliation{Institute of Plasma Physics, CAS, v.v.i., Za Slovankou 3, 182 21 Prague 8, Czech Republic} 

\author{Igor Andriyash}
\affiliation{Department of Physics of Complex Systems, Weizmann Institute of Science, Rehovot 7610001, Israel}

\author{Julien Gautier}
\affiliation{Laboratoire d'Optique Appliqu\'ee, ENSTA, CNRS UMR7639, Ecole Polytechnique, Chemin de la Huni\`ere,91761 Palaiseau, France.}

\author{Stephane Sebban}
\affiliation{Laboratoire d'Optique Appliqu\'ee, ENSTA, CNRS UMR7639, Ecole Polytechnique, Chemin de la Huni\`ere,91761 Palaiseau, France.}

\author{Slava Smartsev}
\affiliation{Department of Physics of Complex Systems, Weizmann Institute of Science, Rehovot 7610001, Israel}

\author{Noemie Jourdain}
\author{Uddhab Chulagain}
\affiliation{ELI Beamlines, Dolni Brezani, Czech Republic}

\author{Yasmina Azamoum}
\affiliation{Laboratoire d'Optique Appliqu\'ee, ENSTA, CNRS UMR7639, Ecole Polytechnique, Chemin de la Huni\`ere,91761 Palaiseau, France.}

\author{Amar Tafzi}
\author{Jean-Philippe Goddet}
\author{Cedric Thaury}
\author{Antoine Rousse}
\affiliation{Laboratoire d'Optique Appliqu\'ee, ENSTA, CNRS UMR7639, Ecole Polytechnique, Chemin de la Huni\`ere,91761 Palaiseau, France.}

\author{Kim Ta Phuoc}
\email[]{kim.taphuoc@ensta.fr}
\affiliation{Laboratoire d'Optique Appliqu\'ee, ENSTA, CNRS UMR7639, Ecole Polytechnique, Chemin de la Huni\`ere,91761 Palaiseau, France.}

\date{\today}

\begin{abstract}

Betatron x-ray sources from laser-plasma accelerators combine compactness, high peak brightness, femtosecond pulse duration and broadband spectrum. However, when produced with Terawatt class lasers, their energy was so far restricted to a few kilo-electronvolt (keV), limiting the range of possible applications. Here we present a simple method to increase the energy and the flux by an order of magnitude without increasing the laser energy. The orbits of the relativistic electrons emitting the radiation were controlled using density tailored plasmas so that the efficiency of the Betatron source is significantly improved.


\end{abstract}


\maketitle

\section{Introduction}

Betatron x-ray sources \cite{PRL2004Rousse,Corde2013} from laser-plasma interaction have the potential to become invaluable tools to reveal ultrafast dynamics at the atomic scale length. In particular, their broadband spectrum and femtosecond duration are ideal features for femtosecond (fs) x-ray absorption spectroscopy applications \cite{Mahieu2018}. However, they remain marginal in the panel of the commonly used x-ray sources, mainly because of their limited photon energies and relatively low average flux. In this letter, we show that the use of density tailored plasmas can dramatically improve the efficiency of Betatron sources and push their energy and flux in the typical range of conventional synchrotron facilities.

A Betatron source reproduces the principle of a synchrotron radiation in a millimeter scale laser-produced plasma \cite{PRL2004Kiselev,PRL2004Rousse}. An ion cavity created in the wake of an intense femtosecond laser simultaneously acts as an accelerator and a wiggler. Electrons trapped in the cavity are accelerated in the longitudinal direction ($\hat{z}$) and are wiggled in the transverse direction ($\hat{x}$,$\hat{y}$) by strong space-charge electromagnetic fields. When electrons reach relativistic energies, they emit synchrotron-like radiation in the x-ray energy range -- the so-called Betatron radiation. All the features of the emitted radiation depend on the electron orbits which are defined by Lorentz factor $\gamma$ (is $\gg1$) of the electron, its transverse oscillation amplitude $r_\beta$, and the background electron plasma density $n_e$. The oscillation frequency of the electron is given by ${\omega_\beta = \omega_{p}/\sqrt{2\gamma}}$, where ${\omega_{p} =k_{p}c= \sqrt{4\pi c^2 r_e n_e}}$ is a plasma frequency, and $r_e$ is the classical electron radius \cite{Corde2013}. We define the parameter $K= r_\beta k_{p}\sqrt{\gamma/2}$, called the wiggler parameter. For $K\gg 1$, typical for laser-plasma accelerators, the Betatron radiation is emitted into an aperture angle $\theta\approx (1+K)/\gamma$ and has a broad spectrum extending up to a critical frequency $\omega_c=3/2 K \gamma^2 \omega_\beta$ after which it rapidly drops. The effective number of photons produced by each electron per oscillation period can be estimated as $N_{ph} \simeq K/30$. When produced with tens of Terawatts class lasers, the Betatron radiation is emitted by electrons with energies in the hundred MeV range. The source delivers few femtosecond x-ray pulses with a broadband spectrum extending up to a few keV and containing about $10^6$~photons/shot/0.1 BW at 1 keV \cite{PoP2005TaPhuoc,PoP2015Schnell}. 

Several paths to increase the flux and photon energy of the betatron source have been studied. The most straightforward way is to increase the laser power. It results in an increase of the electrons energy \cite{NatPhys2006Leemans, Kim2013, Kim2017} and therefore in the emission of brighter and more energetic radiation \cite{NatPhys2010Kneip, Wang2012, Fazel2016, Wood2018}. However, this comes at the cost of a lower repetition rate, inherent to the large scale lasers, which is unattractive for the applications. Alternatively, one promising option is to tailor the plasma density profile in order to control the electron orbits. In reference \cite{TaPhuoc2008}, longitudinal density tailoring was studied theoretically and numerically. It was shown that $r_\beta$, $\gamma$ and $\omega_\beta$ can be increased for appropriately chosen density profiles. This results in a drastically improved efficiency of the Betatron source. Here we present the experimental study of the Betatron x-ray radiation in plasmas with the controllable  longitudinal and transverse density gradients. We show that $\omega_c$ and the integrated radiated energy can be increased by an order of magnitude as compared with the commonly used constant density plasma, referred to as the reference case.

\section{Radiation enhancement with tailored density profiles}

\begin{figure}
\includegraphics[width=0.9 \linewidth]{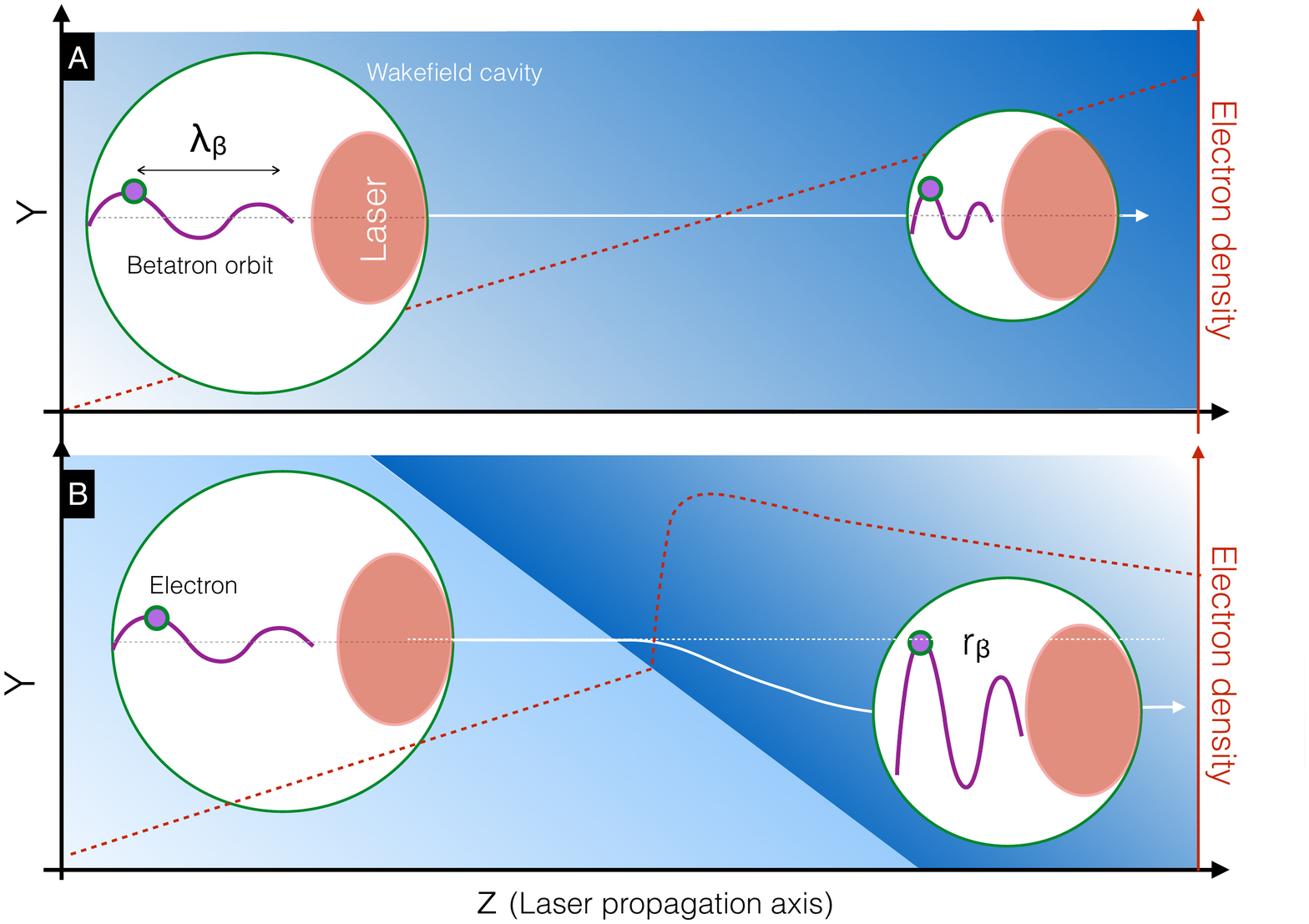} 
\caption{Schematic representation of the density gradients used to improve the efficiency of the Betatron source. A - A longitudinal density gradient contributes to electron re-phasing and reduction of the Betatron period. B - A sharp transverse density gradient results in re-phasing, reduction of the Betatron period and a shift of the cavity axis (an increase of the oscillation amplitude).}
\label{fig1}
\end{figure}

\Cref{fig1} schematically shows how density tailored plasmas can be used to modify the orbits of the electrons oscillating in the cavity. Two interaction scenarii, that can be realized experimentally, are presented. In \cref{fig1}a, plasma density profile has a longitudinal density up-ramp along the laser propagation ($\hat{z}$-axis). As laser pulse travels through the plasma, the wakefield amplitude grows, hence, the electron oscillation frequency $\omega_\beta$ increases and the plasma ion cavity shrinks (the cavity radius is $\propto n_e^{-1/2}$). Shrinking the cavity counteracts the de-phasing which occurs when particles start to overrun the plasma wake. Electrons are therefore maintained in the strongest field region \cite{Rittershofer2014,Guillaume2015} and reach a higher energy as compared to the reference case. With $\gamma$ and $\omega_\beta$ being increased, the Betatron radiation is expected to become brighter and more energetic.

A density gradient along the transverse (e.g. along $\hat{y}$) direction provides another degree of optimization. \Cref{fig1}b represents the case where a tilted density feature refracts the laser pulse and the associated plasma wake. When laser traverses the ascending gradient its axis is deviated, and in the descending gradient the deviation direction reverses. For a sufficiently sharp gradient, $l_\text{grad}\lesssim\lambda_\beta=2\pi c/\omega_\beta$, the electrons oscillation amplitude $r_\beta$ is increased by a quantity equal to the shift of the laser axis \cite{Yu2018}, thus leading to the emission of more energetic and brighter x-rays without angular deviation. Both longitudinal and transverse density gradients can be combined to further enhance the efficiency of the Betatron source \cite{TaPhuoc2008}.

\section{Results}

The experiment has been performed at Laboratoire d'Optique Appliquée using a 50 TW, 30 fs, linearly polarized (along $\hat{x}$-axis) laser, focussed with a $f/20$ parabolic mirror onto a gas target containing a mixture of helium (99~$\%$) and nitrogen ($1 \%$) gases. Accelerated electrons were bent towards a scintillator screen using the static magnet (1 Tesla field over 40~cm). However, in our parameter range ($n_e>10^{19}$~cm$^{-3}$, gas length $\sim 5$~mm) the propagation length exceeds the de-phasing length, and measured spectra are not representative of the energy of the electrons when they emit most of the Betatron radiation. X-rays were observed using either a deep-depletion x-ray CCD (for radiation up to 15 keV) or a scintillator screen imaged with a 16-bit camera (for the radiation up to 100~keV). Pairs of Ross filters were used to characterize the radiation in the range from 2 to 80~keV. Plasma density profiles were estimated using a Normarsky interferometer. A second laser pulse (300~mJ~/~30~fs) was used as a machining beam to estimate where electrons are injected and where x-ray radiation is produced along the laser propagation axis \cite{Thaury2013,Pai:PoP2005}.

\subsection{Slow longitudinal ramp}

We first studied the case of a slow longitudinal gradient. For this, we compared the Betatron radiation from two nozzles: one with a constant density profile and the other with an up-ramp density profile. The density measurements are shown in \cref{fig3}a. In both cases the x-ray emission was maximized by adjusting the nozzle position with respect to the laser focus and the gas pressure. The x-ray signal was measured through an array of Aluminium, Copper and Titanium filters. The Betatron spectra that best fits the measured x-ray signals are represented by the shaded areas in \cref{fig4}. As expected the x-ray signal is improved significantly. The critical energy shifts from $\simeq 5$ to $\simeq 10$ keV and the flux is enhanced by a factor $\simeq 3$. In order to verify, that the signal enhancement results from the change of electron orbits associated with the up-ramp density, we have estimated the x-ray emission regions using the method described in \cite{Thaury2013}. We have found that the plasma lengths, over which Betatron radiation is produced, were $1.5 \pm 0.5$~mm for the up-ramp density, and $ 2 \pm 0.5$~mm for the constant density cases. Such difference of the propagation lengths cannot account for the observed signal enhancement, which confirms that the density gradient itself has the major effect.
This result is confirmed by test-particle simulations based on the ideal ion cavity model \cite{Phuoc2005}. With this simplified approach we have identified the basic parameters which fit the Betatron radiation features \cite{Corde2011}. In \cref{fig4}, the fit of the experimental spectrum for the constant density case is obtained using as initial conditions $r_\beta = 1.25$ $\mu$m, $n_e=10^{19}$~cm$^{-3}$ and a 2~mm propagation distance (this choice of parameters is not exclusive but this has no importance since we focus on the relative differences in the spectra calculated with and without gradient). The spectrum for the up-ramp is well reproduced using the same initial conditions, but with the density profile that increases from $10^{19}$~cm$^{-3}$ to $2\times10^{19}$~cm$^{-3}$ over 2~mm.


\subsection{Sharp tilted ramp}

\begin{figure*}[ht!]
\includegraphics[height=0.8\textheight]{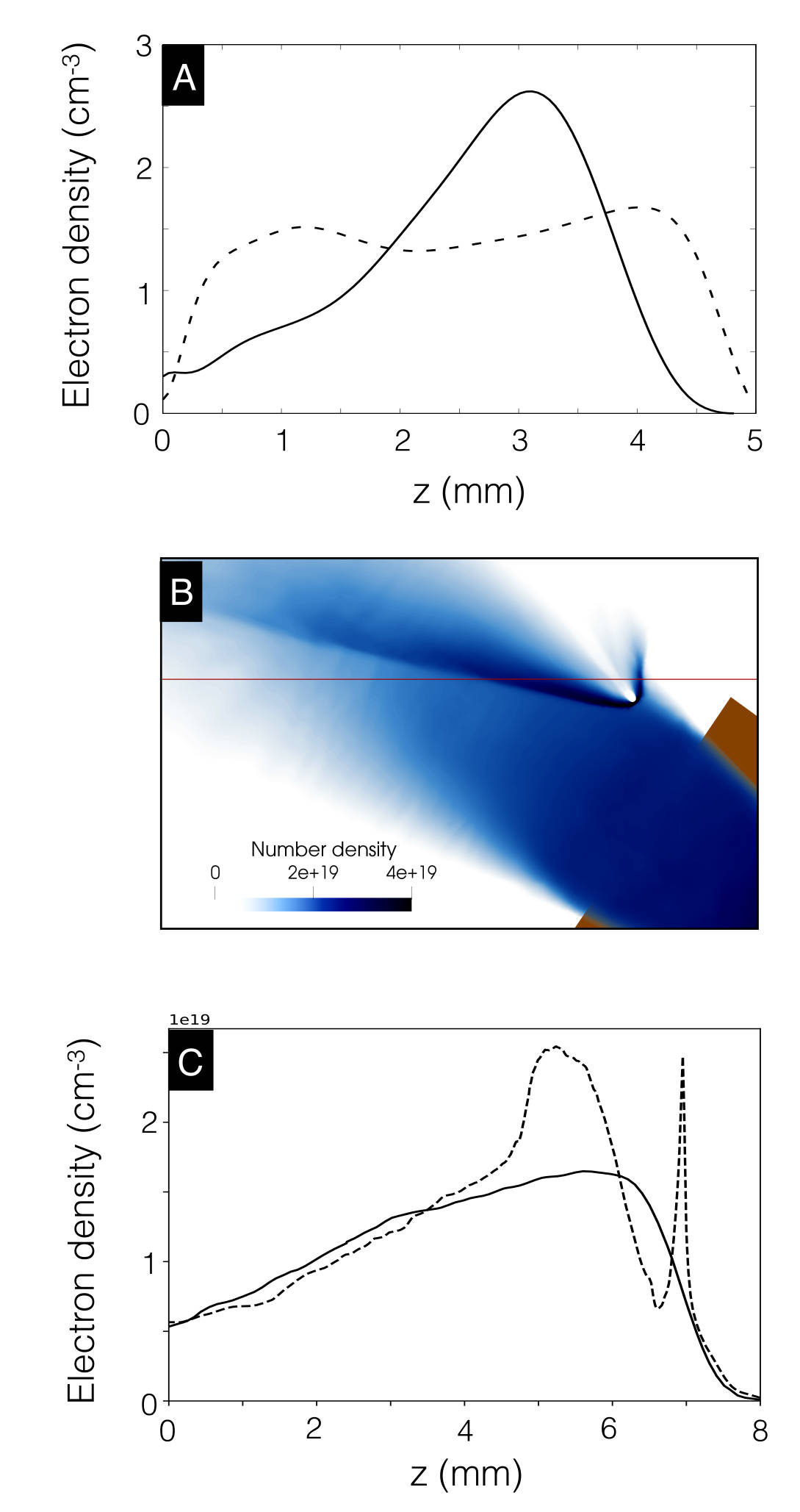} 
\caption{A- Density profiles of the nozzles used for the reference case (dotted line) and the longitudinal gradient (solid line). Beveled or a tilted gas jet were used to produce the longitudinal gradient. B and C- Openfoam simulation of the  transverse gradient was obtained by placing a 100 microns wire can be placed on top of the nozzle. Density profile along the laser pulse propagation (solid line) and reference density profile (dotted line). The wire produces both a sharp longitudinal and transverse density gradient.}\label{fig3}
\end{figure*}

\begin{figure*}
\includegraphics[width=0.9\linewidth]{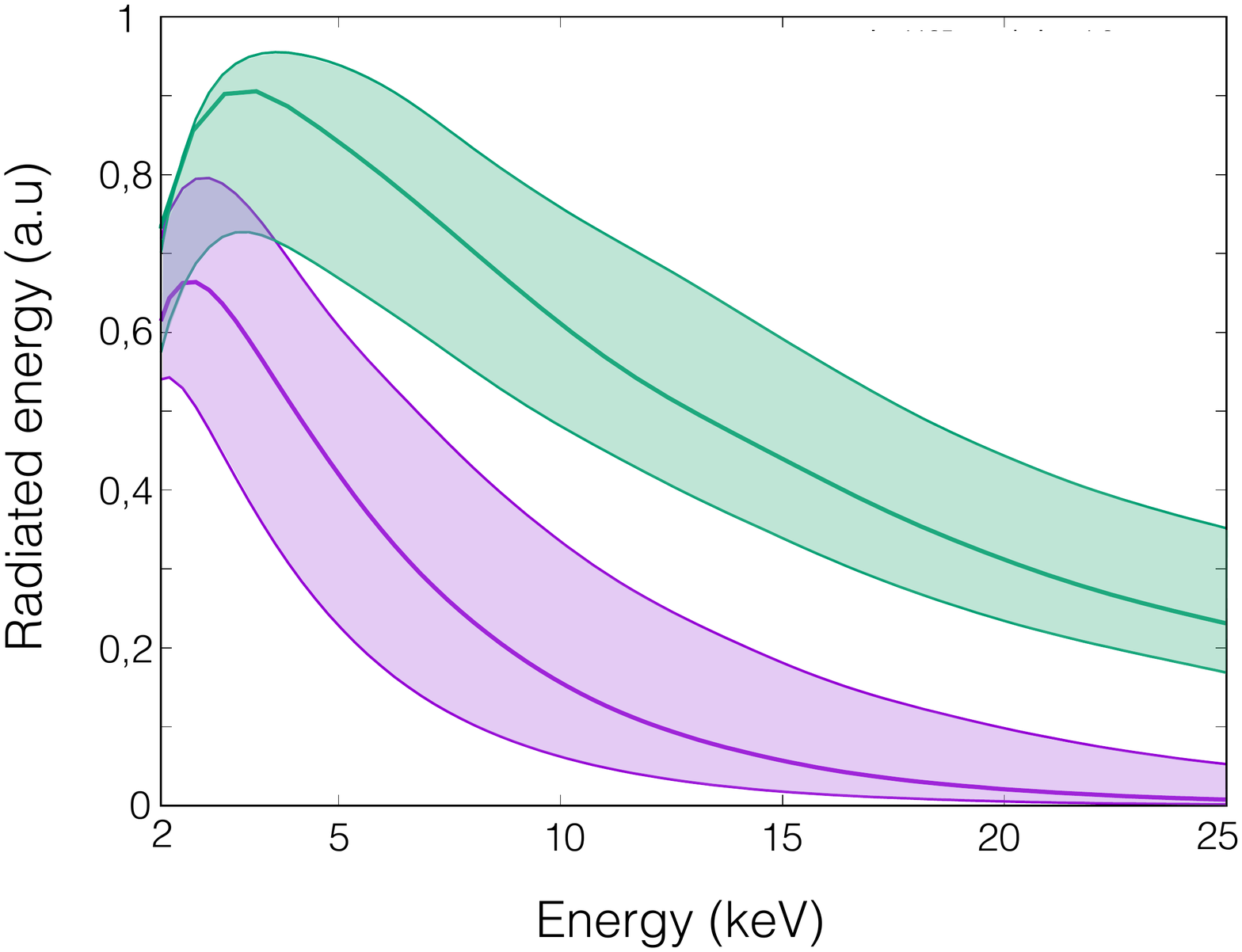}
\caption{Betatron radiation spectra for a constant density plasma (purple) and an up ramp density plasma (green). The shaded area represents the interval of confidence of the measurements. The thick solid lines within these areas are the fits obtained from particle-test simulations. Only the density gradient, set as initial condition, distinguishes the two cases.}\label{fig4}
\end{figure*}

The effect of the sharp tilted transverse density gradient was then studied. It was created by inserting a $100$~$\mu$m diameter wire into the gas flow from a tilted supersonic nozzle. For an accurate estimate of the density profile we performed two-dimensional gas flow modeling using OpenFOAM software \cite{OpenFOAM:2009}. The result is shown in \cref{fig3}b and \cref{fig3}c. 

\begin{figure*}
\includegraphics[width=0.9\linewidth]{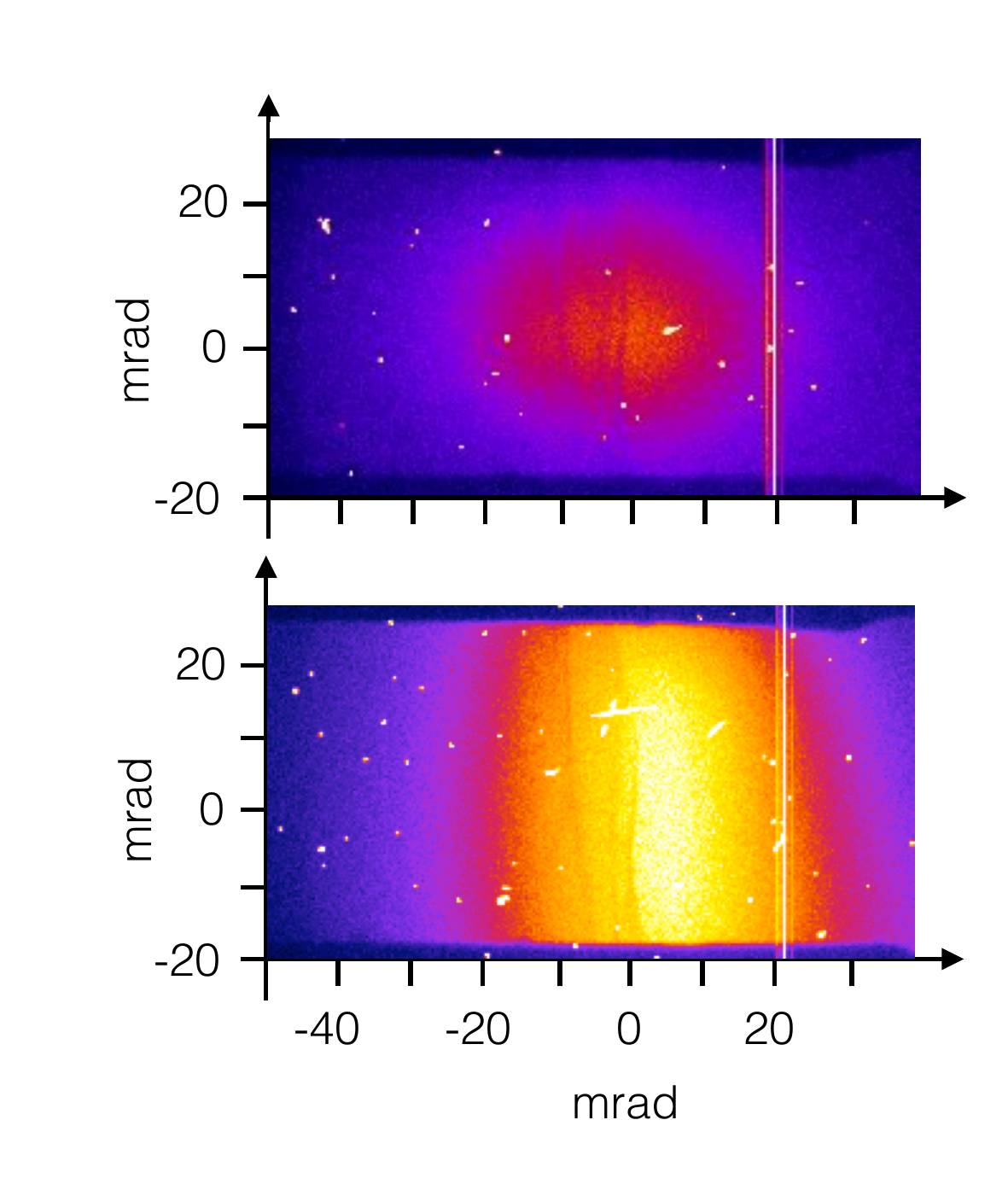} 
\caption{Measured x-ray beam profiles obtained using a scintillator screen (the color scale is identical for both figures). Upper part corresponds to the case without the wire and lower part to the case with the wire. (The vertical line is a damage on the camera)}\label{fig5}
\end{figure*}

The x-ray radiation was characterized with and without the wire. \Cref{fig5} presents typical x-ray beam profiles. Without the wire, the x-ray beam is quasi-circular with a mean divergence $\theta_r\simeq 20 \pm 2$~mrad. When the wire is inserted, the radiation profile becomes elliptical with a main axis in the $\hat{y}$-axis. The mean divergences are ${\theta_x=13 \pm 2}$~mrad and ${\theta_y=29 \pm 3}$~mrad.

\begin{figure}[!ht]
\includegraphics[width=0.9\linewidth]{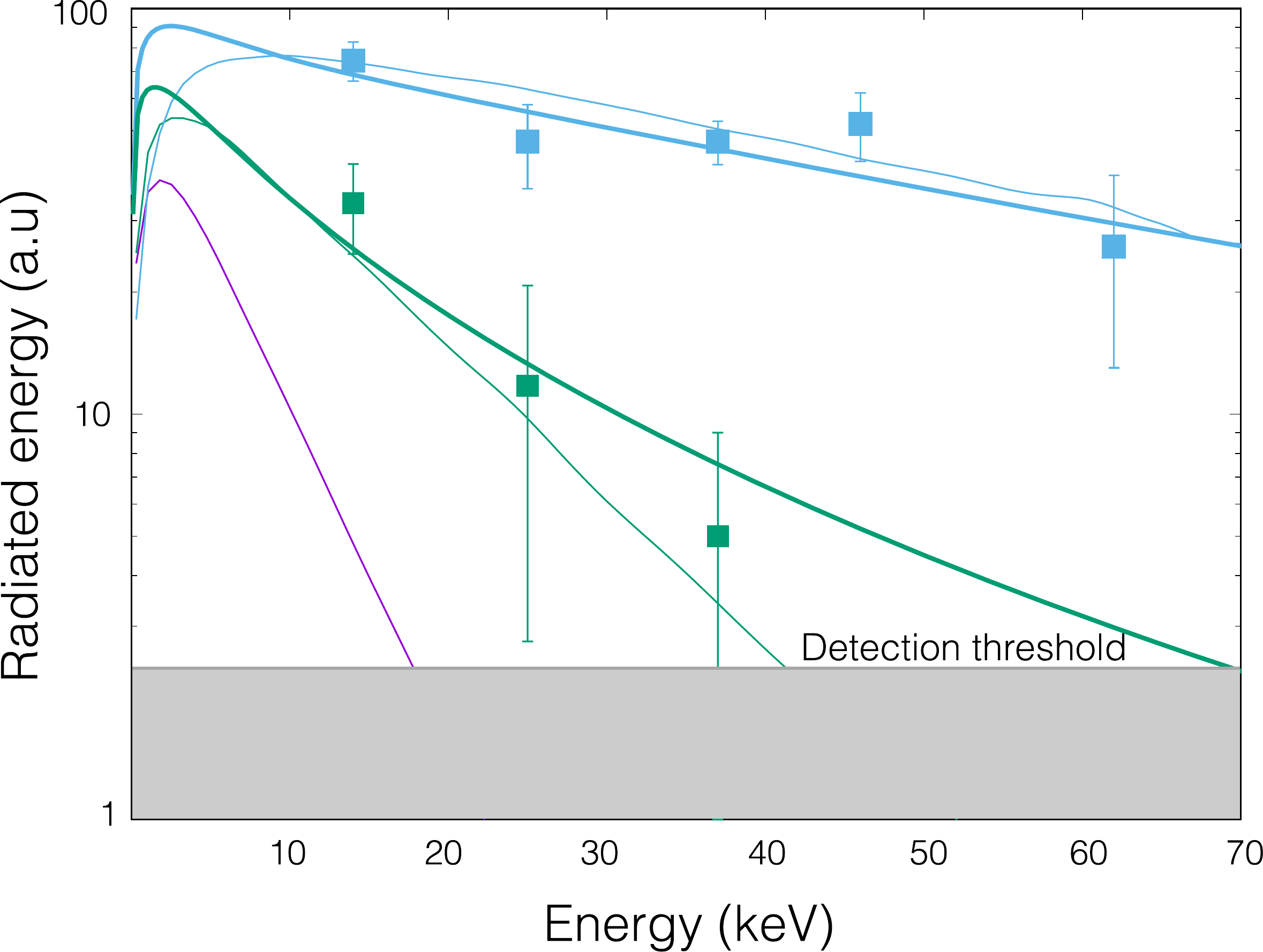}
\caption{Measured Betatron radiation spectra and numerical fits from test-particle and PIC simulations. Thin and thick lines respectively correspond to test-particle and PIC simulations. Purple line is for the reference case. Green is for the up-ramp case. Blue is for the up-ramp and wire.}\label{fig6}
\end{figure}

The analysis of the x-ray beam profiles provides useful information as there is a direct correlation between x-ray beam profile and electrons orbits \cite{PRL2006TaPhuoc}. In particular, the decrease of the x-ray beam divergence along the $\hat{x}$-axis (perpendicular to the shock tilt) is a signature of an increase of $\gamma$. From the divergence measurement, we could estimate an electron energy gain of the order of 40-50~$\%$ assuming the oscillation amplitude along the $\hat{x}$-axis, $r_{\beta_x}$, constant. In the direction of the density gradient tilt, the radiation divergence is significantly increased. This is the consequence of an increase of the electron oscillation amplitude $r_{\beta_y}$ along the $\hat{y}$-axis. The measured asymmetry with the wire translates to the ratio of the oscillation amplitudes $\theta_y/\theta_x = r_{\beta_y}/r_{\beta_x} \simeq 2.2$. 
These deductions confirm that the transverse density gradient produced by the shock allows to increase both $\gamma$ and $r_\beta$. The spectra with and without wire were measured using Ross filter pairs. They are presented in \cref{fig6} together with the reference spectrum. The thin solid lines correspond to the spectra obtained using the test-particle simulations with the same parameters as before, but with $r_\beta = 2.5$ in the case of the wire, which is in good agreement with the ratio of $r_{\beta_y}/r_{\beta_x}$ deduced from the x-ray beam profiles. Fitting the obtained data with the standard synchrotron spectra, we can estimate the critical energies at $50$~keV and $10$ keV with and without the wire respectively, and at $5$ keV for the reference case. The total radiated x-ray energy is further increased 2.5 times when using the wire. From a series of systematic shots, we have found, that the effect was sensitive to the wire $\hat{z}$-position within a $\pm 500$~$\mu$m interval, and the shock is required to be sharp to the level of typically a hundred of microns. Optimizing wire position ensured that electrons are trapped and accelerated to the maximum energies before the shock.

\subsection{Particle-in-Cell simulations}

For an additional insight into the physics of x-rays enhancement, we performed particle-in-cell (PIC) simulations. We have used the quasi-3D pseudo-spectral code FBPIC \cite{Lehe:CPC2016}. The target was considered to be a fully ionized He plasma with 2\% of N$^{+5}$ ion plasma. We considered the cases of gas flow without and with the tilted shock (gray colors in \cref{fig4}). The density profile is an asymmetric Gaussian density profile defined by $n(z<0) = n_0 \exp(-z^2/L_\mathrm{l})$, and $n(z>0) = n_0 \exp(-z^2/L_\mathrm{r})$, where $L_\mathrm{l}=3$~mm and $L_\mathrm{r}=1$~mm. The peak density without wire insertion is $n_0=1.64\times 10^{19}$~cm$^{-3}$. A shock with a peak density $n_0/2$ is added at $z_s=-1.2$~mm; it has an asymmetric Gaussian density profile with $L_\mathrm{sl}=0.1$~mm, $L_\mathrm{sr}=0.7$~mm. The shock tilt angle estimated from CFD simulations is $\theta_s=20^\circ$, and the tilt introduced by replacing $z \to z - z_s + y\tan\theta_s$.

\begin{figure}[!ht]
\includegraphics[width=0.9\linewidth]{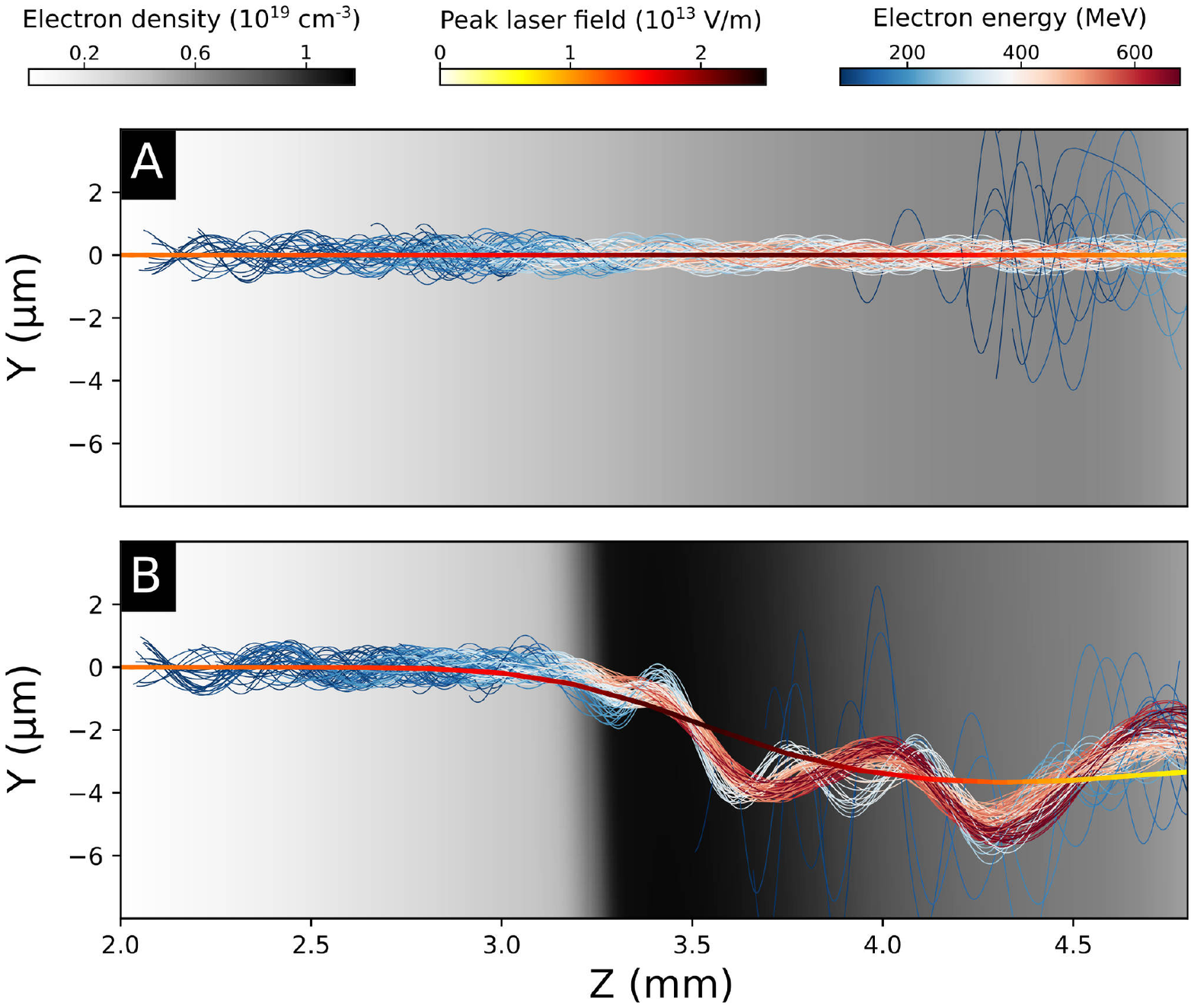}
\caption{Laser propagation (thick curve) and particle trajectories (thin curves) in $(z,y)$-plane colored according to the laser peak field and the particle energy respectively. Gray levels represents plasma density. A - Up-ramp gradient. B - Up ramp gradient and tilted shock.
}\label{fig7}
\end{figure}

\Cref{fig7} shows the plasma density in gray scale in the $(z,y)$-plane, as well as the laser centroid (thick curve) and particles (thin curves) trajectories, colored according to the laser peak field and the particle energy respectively. When only the longitudinal gradient is considered (see \cref{fig7}a), the acceleration is continuous, and the oscillations amplitudes do not change significantly during the interaction. When the shock is added in \cref{fig7}b, the tilt produces laser refraction and leads to the displacement of the propagation axis by $\approx4.5$~$\mu$m. It induces a kick onto accelerated electrons increasing their amplitude of oscillation $r_\beta$. Moreover, the sharp rise of the plasma density at the shock relocates particles to higher accelerating and focussing fields, which boosts electron energies and induces higher frequency oscillations. The spectra for each case were calculated to estimate the overall effect on the betatron emission. They are shown with the thick solid curves in \cref{fig6}. A good agreement with the experimental measurements in the photon energy distributions (blue and green curves) is obtained. The total energy produced per charge is increased by a factor 3, which is close to the experimental values.

\section{Conclusion}

In conclusion we have demonstrated that the efficiency of Betatron sources can be significantly improved by using longitudinal and vertical density gradients. The radiation produced has a critical energy ten times higher than the Betatron radiation produced in an homogeneous plasma. We anticipate that this progress will represent a significant milestone in the development of table top femtosecond x-ray sources. 

*Corresponding author. Electronic address: kim.taphuoc@ensta.fr\\

Acknowledgments\\

MK and UC would like to acknowledge the project ADONIS (CZ 02.1.01/0.0/0.0/16-019/0000789) from ERDF and the Project LQ1606 obtained with the financial support of the Ministry of Education, Youth and Sports as part of targeted support from the National Programme of Sustainability II.

\bibliography{Mabiblio}

\begin{thebibliography}{10}

\bibitem{PRL2004Rousse}
A~Rousse, K~{Ta Phuoc}, R~Shah, A~Pukhov, E~Lefebvre, V~Malka, S~Kiselev,
  F~Burgy, J~P. Rousseau, D~Umstadter, and D~Hulin.
\newblock {Production of a keV X-Ray Beam from Synchrotron Radiation in
  Relativistic Laser-Plasma Interaction}.
\newblock {\em Phys. Rev. Lett.}, 93(13):135005, sep 2004.

\bibitem{Corde2013}
S.~Corde, K.~{Ta Phuoc}, G.~Lambert, R.~Fitour, V.~Malka, A.~Rousse, A.~Beck,
  and E.~Lefebvre.
\newblock {Femtosecond x rays from laser-plasma accelerators}.
\newblock {\em Reviews of Modern Physics}, 85(1), 2013.

\bibitem{Mahieu2018}
B.~Mahieu, N.~Jourdain, K.~{Ta Phuoc}, F.~Dorchies, J.~P. Goddet, A.~Lifschitz,
  P.~Renaudin, and L.~Lecherbourg.
\newblock {Probing warm dense matter using femtosecond X-ray absorption
  spectroscopy with a laser-produced betatron source}.
\newblock {\em Nature Communications}, 9(1):2--7, 2018.

\bibitem{PRL2004Kiselev}
S~Kiselev, A~Pukhov, and I~Kostyukov.
\newblock {X-ray Generation in Strongly Nonlinear Plasma Waves}.
\newblock {\em Phys. Rev. Lett.}, 93(13):135004, 2004.

\bibitem{PoP2005TaPhuoc}
K~{Ta Phuoc}, F~Burgy, J~P Rousseau, V~Malka, A~Rousse, R~Shah, D~Umstadter,
  A~Pukhov, and S~Kiselev.
\newblock {Laser based synchrotron radiation}.
\newblock {\em Phys. Plasmas}, 12(2):23101, 2005.

\bibitem{PoP2015Schnell}
Michael Schnell, Alexander Sävert, Ingo Uschmann, Oliver Jansen,
  Malte~Christoph Kaluza, and Christian Spielmann.
\newblock Characterization and application of hard x-ray betatron radiation
  generated by relativistic electrons from a laser-wakefield accelerator.
\newblock {\em Journal of Plasma Physics}, 81(4):475810401, 2015.

\bibitem{NatPhys2006Leemans}
W~P Leemans, B~Nagler, A~J Gonsalves, C~Toth, K~Nakamura, C~G~R Geddes,
  E~Esarey, C~B Schroeder, and S~M Hooker.
\newblock {GeV electron beams from a centimetre-scale accelerator}.
\newblock {\em Nat. Phys.}, 2(10):696--699, 2006.

\bibitem{Kim2013}
Hyung~Taek Kim, Ki~Hong Pae, Hyuk~Jin Cha, I.~Jong Kim, Tae~Jun Yu, Jae~Hee
  Sung, Seong~Ku Lee, Tae~Moon Jeong, and Jongmin Lee.
\newblock {Enhancement of electron energy to the multi-gev regime by a
  dual-stage laser-wakefield accelerator pumped by petawatt laser pulses}.
\newblock {\em Physical Review Letters}, 111(16):1--5, 2013.

\bibitem{Kim2017}
Hyung~Taek Kim, V.~B. Pathak, Ki~{Hong Pae}, A.~Lifschitz, F.~Sylla, Jung~Hun
  Shin, C.~Hojbota, Seong~Ku Lee, Jae~Hee Sung, Hwang~Woon Lee, E.~Guillaume,
  C.~Thaury, Kazuhisa Nakajima, J.~Vieira, L.~O. Silva, V.~Malka, and Chang~Hee
  Nam.
\newblock {Stable multi-GeV electron accelerator driven by waveform-controlled
  PW laser pulses}.
\newblock {\em Scientific Reports}, 7(1):1--8, 2017.

\bibitem{NatPhys2010Kneip}
S~Kneip, C~McGuffey, J~L Martins, S~F Martins, C~Bellei, V~Chvykov, F~Dollar,
  R~Fonseca, C~Huntington, G~Kalintchenko, A~Maksimchuk, S~P~D Mangles,
  T~Matsuoka, S~R Nagel, C~A~J Palmer, J~Schreiber, K~Ta Phuoc, A~G~R Thomas,
  V~Yanovsky, L~O Silva, K~Krushelnick, and Z~Najmudin.
\newblock {Bright spatially coherent synchrotron X-rays from a table-top
  source}.
\newblock {\em Nat. Phys.}, 6(12):980--983, 2010.

\bibitem{Wang2012}
X.~Wang, R.~Zgadzaj, N.~Fazel, S.~A. Yi, X.~Zhang, W.~Henderson, Y.~Y. Chang,
  R.~Korzekwa, H.~E. Tsai, C.~H. Pai, Z.~Li, H.~Quevedo, G.~Dyer, E.~Gaul,
  M.~Martinez, A.~Bernstein, T.~Borger, M.~Spinks, M.~Donovan, S.~Y. Kalmykov,
  V.~Khudik, G.~Shvets, T.~Ditmire, and M.~C. Downer.
\newblock {Petawatt-laser-driven wakefield acceleration of electrons to 2 GeV
  in 1017 cm-3 plasma}.
\newblock {\em AIP Conference Proceedings}, 1507:341--344, 2012.

\bibitem{Fazel2016}
Neil Fazel, Xiaoming Wang, Rafal Zgadzaj, Z.~Li, Xi~Zhang, W.~Henderson,
  H.~Quevedo, G.~Dyer, E.~Gaul, M.~Martinez, M.~Spinks, M.~Donovan, V.~Khudik,
  G.~Shvets, T.~Ditmire, and M.~C. Downer.
\newblock {Betatron x-rays from GeV laser-plasma-accelerated electrons}.
\newblock {\em AIP Conference Proceedings}, 1777(October 2016):1--6, 2016.

\bibitem{Wood2018}
J.~C. Wood, D.~J. Chapman, K.~Poder, N.~C. Lopes, M.~E. Rutherford, T.~G.
  White, F.~Albert, K.~T. Behm, N.~Booth, J.~S.J. Bryant, P.~S. Foster,
  S.~Glenzer, E.~Hill, K.~Krushelnick, Z.~Najmudin, B.~B. Pollock, S.~Rose,
  W.~Schumaker, R.~H.H. Scott, M.~Sherlock, A.~G.R. Thomas, Z.~Zhao, D.~E.
  Eakins, and S.~P.D. Mangles.
\newblock {Ultrafast Imaging of Laser Driven Shock Waves using Betatron X-rays
  from a Laser Wakefield Accelerator}.
\newblock {\em Scientific Reports}, 8(1):1--10, 2018.

\bibitem{TaPhuoc2008}
K.~{Ta Phuoc}, E.~Esarey, V.~Leurent, E.~Cormier-Michel, C.~G.R. Geddes, C.~B.
  Schroeder, A.~Rousse, and W.~P. Leemans.
\newblock {Betatron radiation from density tailored plasmas}.
\newblock {\em Physics of Plasmas}, 15(6):1--10, 2008.

\bibitem{Rittershofer2014}
W~Rittershofer, C~B Schroeder, E~Esarey, F~J Gr{\"{u}}ner, W~P Leemans,
  W~Rittershofer, C~B Schroeder, E~Esarey, F~J Gr{\"{u}}ner, and W~P Leemans.
\newblock {Tapered plasma channels to phase-lock accelerating and focusing
  forces in laser- plasma accelerators Tapered plasma channels to phase-lock
  accelerating and focusing forces in laser-plasma accelerators}.
\newblock 063104(2010), 2014.

\bibitem{Guillaume2015}
E.~Guillaume, A.~D{\"{o}}pp, C.~Thaury, K.~{Ta Phuoc}, A.~Lifschitz,
  G.~Grittani, J.~P. Goddet, A.~Tafzi, S.~W. Chou, L.~Veisz, and V.~Malka.
\newblock {Electron Rephasing in a Laser-Wakefield Accelerator}.
\newblock {\em Physical Review Letters}, 115(15), 2015.

\bibitem{Yu2018}
Changhai Yu, Jiansheng Liu, Wentao Wang, Wentao Li, Rong Qi, Zhijun Zhang,
  Zhiyong Qin, Jiaqi Liu, Ming Fang, Ke~Feng, Ying Wu, Lintong Ke, Yu~Chen,
  Cheng Wang, Yi~Xu, Yuxin Leng, Changquan Xia, Ruxin Li, and Zhizhan Xu.
\newblock {Enhanced betatron radiation by steering a laser-driven plasma
  wakefield with a tilted shock front}.
\newblock {\em Applied Physics Letters}, 112(13), 2018.

\bibitem{Thaury2013}
C.~Thaury, K.~{Ta Phuoc}, S.~Corde, P.~Brijesh, G.~Lambert, S.~P.D. Mangles,
  M.~S. Bloom, S.~Kneip, and V.~Malka.
\newblock {Probing electron acceleration and x-ray emission in laser-plasma
  accelerators}.
\newblock {\em Physics of Plasmas}, 20(6):1--5, 2013.

\bibitem{Pai:PoP2005}
C.-H. Pai, S.-Y. Huang, C.-C. Kuo, M.-W. Lin, J.~Wang, S.-Y. Chen, C.-H. Lee,
  and J.-Y. Lin.
\newblock Fabrication of spatial transient-density structures as high-field
  plasma photonic devices.
\newblock {\em Physics of Plasmas}, 12(7):070707, 2005.

\bibitem{Phuoc2005}
Kim~Ta Phuoc, Fr{\'{e}}deric Burgy, Jean-Philippe Rousseau, Victor Malka,
  Antoine Rousse, Rahul Shah, Donald Umstadter, Alexander Pukhov, and Sergei
  Kiselev.
\newblock {Laser based synchrotron radiation}.
\newblock {\em Physics of Plasmas}, 12(2):023101, 2005.

\bibitem{Corde2011}
S.~Corde, K.~Ta Phuoc, R.~Fitour, J.~Faure, A.~Tafzi, J.~P. Goddet, V.~Malka,
  and A.~Rousse.
\newblock {Controlled betatron X-ray radiation from tunable optically injected
  electrons}.
\newblock {\em Physical Review Letters}, 107(25):2--6, 2011.

\bibitem{OpenFOAM:2009}
Hrvoje Jasak.
\newblock Openfoam: Open source cfd in research and industry.
\newblock {\em International Journal of Naval Architecture and Ocean
  Engineering}, 1(2):89 -- 94, 2009.

\bibitem{PRL2006TaPhuoc}
K~{Ta Phuoc}, S~Corde, R~Shah, F~Albert, R~Fitour, J~P. Rousseau, F~Burgy,
  B~Mercier, and A~Rousse.
\newblock {Imaging Electron Trajectories in a Laser-Wakefield Cavity Using
  Betatron X-Ray Radiation}.
\newblock {\em Phys. Rev. Lett.}, 97(22):225002, 2006.

\bibitem{Lehe:CPC2016}
Rémi Lehe, Manuel Kirchen, Igor~A. Andriyash, Brendan~B. Godfrey, and Jean-Luc
  Vay.
\newblock A spectral, quasi-cylindrical and dispersion-free particle-in-cell
  algorithm.
\newblock {\em Computer Physics Communications}, 203:66 -- 82, 2016.

\end{thebibliography}
\bibliographystyle{unsrt}

\newpage

\end{document}